\documentclass[aps,prl,twocolumn,epsf,epsfig,amsmath]{revtex4}

\usepackage{latexsym}
\usepackage{hyperref}
\usepackage{times}
\usepackage{graphicx}
\usepackage{amsmath}
\usepackage{multirow}

\usepackage{dcolumn}
\usepackage{bm}
\usepackage{ textcomp }
\usepackage{color}
\usepackage{amssymb}
\usepackage{xcolor}
\usepackage{easyReview}
\usepackage{mathdots}
\usepackage{float}
\usepackage{lineno}
\usepackage{todonotes}
\usepackage{array}
\newcommand{\beq}{\begin{eqnarray}}
\newcommand{\eeq}{\end{eqnarray}}

\newcommand{\disp}[1]{Eq. (\ref{#1})}

\newcommand{\refdisp}[1]{Ref. [\onlinecite{#1}]}
\newcommand{\figdisp}[1]{Fig. \ref{#1}}

\begin{document}
\title{Topological Phase Transition without Single-Particle-Gap Closing in Strongly Correlated Systems}
\author{ Peizhi Mai$^{1}$, Jinchao Zhao$^{1}$, Thomas A. Maier$^{2}$, Barry Bradlyn$^{1}$
 and Philip W. Phillips$^{1,\dagger}$ }

\affiliation{$^1$Department of Physics and Institute of Condensed Matter Theory, University of Illinois at Urbana-Champaign, Urbana, IL 61801, USA}
\affiliation{$^2$Computational Sciences and Engineering Division, Oak Ridge National Laboratory, Oak Ridge, Tennessee 37831, USA}


\begin{abstract}
We show here that numerous examples abound where changing topology does not necessarily close the bulk insulating charge gap as demanded in the standard non-interacting picture. From extensive determinantal and dynamical cluster quantum Monte Carlo simulations of the half-filled and quarter-filled Kane-Mele-Hubbard model, we show that for sufficiently strong interactions at either half- or quarter-filling, a transition between topological and trivial insulators occurs without the closing of a charge gap.  To shed light on this behavior, we illustrate that an exactly solvable model reveals that while the single-particle gap remains, the many-body gap does in fact close. These two gaps are the same in the non-interacting system but depart from each other as the interaction turns on. We purport that for interacting systems, the proper probe of topological phase transitions is the closing of the many-body rather than the single-particle gap.  
\end{abstract}
\date{October 2023}


\maketitle

\section{Introduction}
The discovery of topological insulators\cite{Haldane,bhz,kanemele1,kanemele2,Fu1,jmoore,rahulroy2} has highlighted the fact that the energy spectrum alone does not fully caputre the behavior of electronic systems. In systems with broken time-reversal symmetry, band inversions driven by complex next-nearest-neighbor hopping---as in the Haldane\cite{Haldane} model---lead to Chern insulating states with nontrivial Berry curvature. Similarly, in time-reversal invariant models such as the Kane-Mele\cite{kanemele1} and Bernevig-Hughes-Zhang\cite{bhz} models, spin-orbit coupling can open a topological gap leading to a quantum spin Hall state with nontrivial spin Chern numbers and $\mathbb{Z}_2$ invariant. In noninteracting systems such as these, the nontrivial topology is encoded in the analytic properties of the occupied single-particle wave functions over the Brillouin zone, which cannot change under adiabatic deformations that maintain an insulating gap without breaking any symmetries. Consequently, a transition from a topologically trivial to a topologically nontrivial insulating state must involve the closing of an energy gap, or the breaking of a symmetry (such as time-reversal invariance) that protects the topological phase. Since in non-interacting systems the energy gap is entirely determined by the single-particle band structure, a topological phase transition (TPT) without symmetry breaking in non-interacting systems also requires the single-particle charge gap to close.



This conventional understanding of a TPT has now been challenged by recent experiments on strongly correlated topological systems. Specifically,  experiments on AB stacked MoTe$_2$/WSe$_2$ moir\'e heterobilayers\cite{TingxinLi} find a quantum spin/valley Hall effect at $\nu=2$ resembling the Kane-Mele physics and the emergence of a ferromagnetic quantum anomalous Hall (QAH) insulator at $\nu=1$ (one hole per moir\'e unit cell) driven by strong correlations. Also, by tuning the gate voltage, they observed a transition to a paramagnetic trivial Mott insulator without closing the charge gap. Interestingly, two subsequent experiments\cite{nogapxiaodong,YihangZeng} on twisted moir\'e bilayer MoTe$_2$ also found a QAH insulator at $\nu=1$ and a similar TPT but signaled by closing the charge gap. Another experiment\cite{foutty2023mapping} on twisted bilayer WSe$_2$ identified a similar transition at $\nu=1$ with substantial suppression of the charge gap though they were not able to determine if a complete gap closing was obtained. Crucially, each of these nontrivial topological phases is known to be realizable in weakly correlated models. Hence, it is naively expected that any transition between them would require either the breaking of the protected symmetry or the closure of a (single-particle) charge gap, as in non-interacting models. Motivated by these paradoxical results, we study here generally how a TPT obtains without closing the single-particle gap in the presence of strong correlations.  

The issue of a TPT in the absence of the closure of the single-particle gap (which we denote as a ``gap-not-closing TPT'') was initially discussed in \refdisp{Nagaosa}. Those authors demonstrated that in a non-interacting system if a TPT is associated with explicit symmetry breaking (driven by a symmetry-breaking field) that affects a change of topological classes, then the gap does not need to close. This same principle applies to the mean-field theory for interaction-driven TPTs in which the Hamiltonian lacks an explicit symmetry-breaking term, but the ground state spontaneously breaks the symmetry. For instance, earlier studies of the Bernevig-Hughes-Zhang-Hubbard model\cite{Amaricciprl2015,Crippaprl2021,Paoletti2023} and the Kane-Mele-Hubbard (KMH) model\cite{Hohenadler2011,Hohenadler2012,WeiWu} primarily examined half-filling (corresponding to $\nu=2$ in the experiment) and observed a phase transition from a quantum spin Hall (QSH) insulator to a trivial Mott insulator without closing the single-particle gap as $U$ increases. By contrast, a recent study\cite{quarter} on the KMH model at quarter-filling (corresponding to $\nu=1$) found the general emergence of a topological Mott insulator in the presence of strong correlations, consistent with the experiments\cite{TingxinLi,nogapxiaodong,YihangZeng,foutty2023mapping}. Whether there is a closing of the single-particle gap in its transition to a trivial Mott insulator is yet to be determined and requires a more accurate treatment on the interactions. Such a transition can not be captured even qualitatively by \refdisp{Nagaosa} since there is no non-interacting counterpart for both insulating states. 



In this paper, we address the conditions under which a TPT driven by strong correlations requires the closing of a single-particle charge gap. We first study the generalized KMH model at half- and quarter-filling using the unbiased determinantal quantum Monte Carlo (DQMC) and the dynamical cluster approximation (DCA). We find that in the presence of strong interactions, a TPT can occur even though the single-particle (charge) gap remains open between a topological insulator and a trivial Mott insulator at both 1/2 and 1/4 filling. This observation challenges our conventional understanding of a TPT. To analyze this situation in more details, we then study the exactly soluble KM model with extended orbital Hatsugai-Kohmoto (HK)\cite{HK,HKnp1,HKnp2,quarter,barry,zhaoprl} interaction which provides an example of such a TPT at half-filling. We find generically that the single-particle gap departs from the many-body gap in the presence of interactions. It is the closure of the many-body gap that indicates the change of topology, while the single-particle gap, usually measured in the transport experiment, could remain open during the transition.  Experiments designed to measure the many-body gap can directly probe this claim. 

\section{Results}

\subsection{Hubbard interaction}
We consider the generalized KMH model in an external magnetic field,
\beq
\begin{aligned}
\label{Eq:KMHubbardBfield}
    H_{\text{KMH}}=&H_{\text{KM}}+H_{\text{Hubbard}},\\
    H_{\text{KM}}=&t\sum_{\langle{\bf i}{\bf j}\rangle\sigma} e^{i \phi_{{\bf i},{\bf j}}} 
    c^\dagger_{{\bf i}\sigma}c^{\phantom\dagger}_{{\bf j}\sigma}+t'\sum_{\langle\langle{\bf i}{\bf j}\rangle\rangle\sigma}e^{\pm i \psi\sigma} e^{i \phi_{{\bf i},{\bf j}} } 
    c^\dagger_{{\bf i}\sigma}c^{\phantom\dagger}_{{\bf j}\sigma} \\& -\mu\sum_{{\bf i},\sigma} n_{{\bf i}\sigma} + \lambda_\nu(\sum_{{\bf i}\in\text{A},\sigma}n_{{\bf i}\sigma}-\sum_{{\bf i}\in\text{B},\sigma}n_{{\bf i}\sigma})
     \\
    H_{\text{Hubbard}}=& U\sum_{{\bf i}}(n_{{\bf i}\uparrow}-\frac{1}{2})(n_{{\bf i}\downarrow}-\frac{1}{2}), \label{KMH}
\end{aligned}
\eeq
where we include the nearest-neighbor hopping $t=1$ between sites on the two different sublattices on the honeycomb lattice as the energy scale. The next-nearest-neighbor hopping $ t^\prime \text{e}^{\pm i\psi\sigma}$ plays the role of spin-orbit coupling with the Haldane phase $\psi$\cite{Haldane} 
, and $\lambda_v$ is the sublattice potential difference. If $\psi=0.5$ (in the unit of $\pi$), the hopping reduces to the original KM model. In the non-interacting case, the two lower bands are in the QSH regime provided $\lambda_v<3\sqrt{3}t'\sin{\psi}$. The phase factor $\exp(i \phi_{{\bf i},{\bf j}})$ resulting from the standard Peierls substitution contains the effect of the external magnetic field, which is introduced to probe the zero-field topology and minimize finite-size effects\cite{mfp,Haldanequarter,quarter}. Here $\phi_{{\bf i},{\bf j}}=(2\pi /\Phi_0) \int_{r_{\bf i}}^{r_{\bf j}} {\bf A}\cdot d{\bf l}$, where $\Phi_0=e/h$ represents the magnetic flux quantum, the vector potential ${\bf A}=(x\hat{y}-y\hat{x})B/2$ (symmetric gauge), and the integration is along a straight-line path. To maintain the single-valueness of the wavefunction in a finite system with the adjusted periodic boundary condition requires the flux quantization condition $\Phi/\Phi_0=n_f/N_c$ with $n_f$ an integer and $N_c$ the number of unit cells.

\begin{figure*}[ht]
    \centering
    \includegraphics[width=0.7\textwidth]{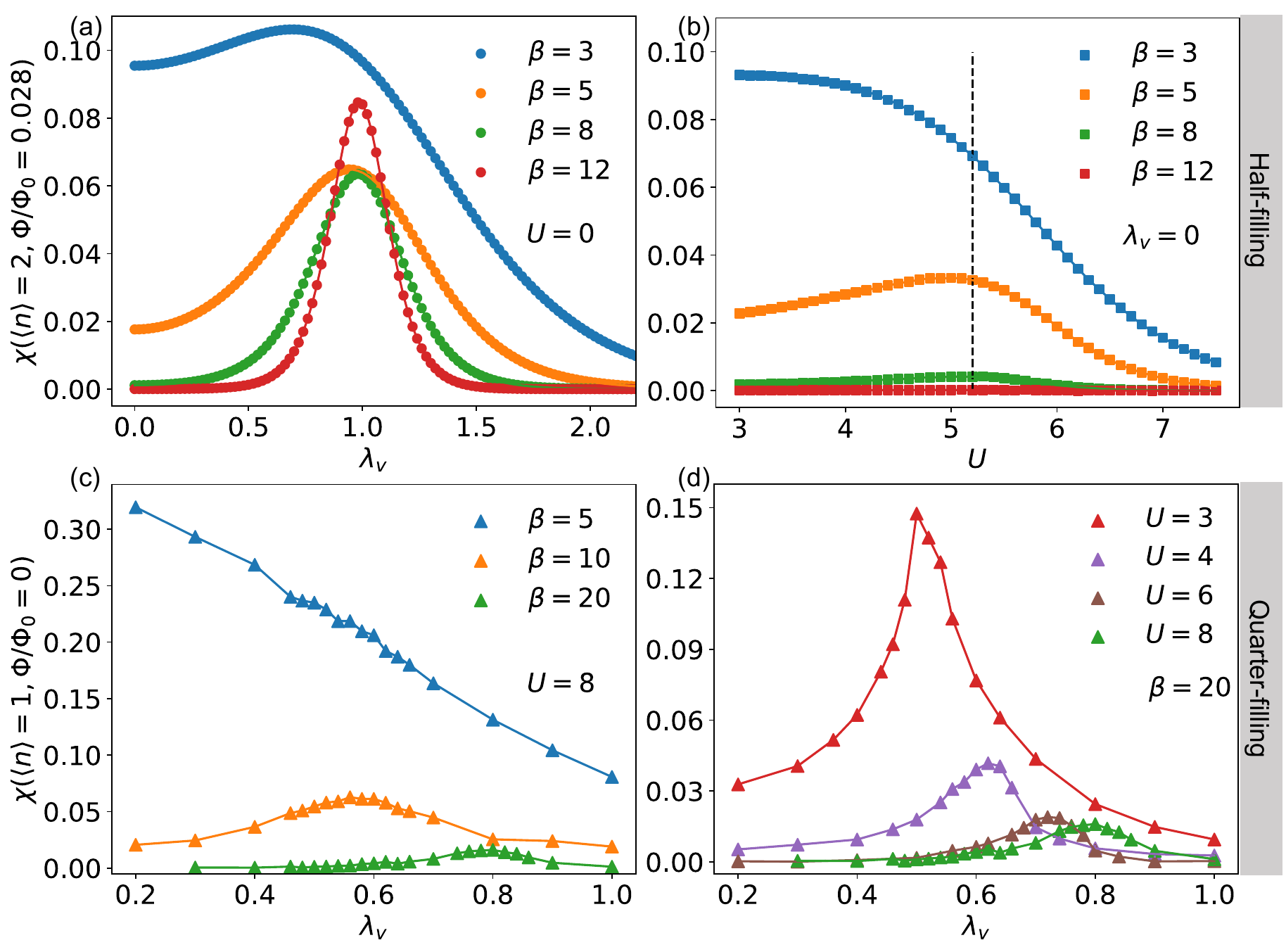}
    \caption{Panel (a) and (b) show the compressibility at the minimal field and half-filling from DQMC simulations as a function of $\lambda_v$ with $U=0$ and as a function of $U$ with $\lambda_v=0$, respectively, for different temperatures. Both indicate a topological phase transition from a QSH insulator to a trivial insulator. $t'/t=0.2$ and $\psi=0.5$ are set for both panels. Panel (c) displays the compressibility at zero field and quarter-filling as a function of $\lambda_v$ with varying temperature while fixing $U=8$.  Panel (d) presents the same quantity but by varying $U$ while fixing $\beta=20$. $t'/t=0.3$ and $\psi=0.81$ are set for panels (c) and (d).
    }
    \label{fig:chi}
\end{figure*}

We simulate this model using DQMC method with jackknife resampling to estimate the error bars. We focus on a lattice with $6\times6$ unit cells; finite size effects are suppressed considerably with a small magnetic flux\cite{mfp}. To access the charge gap evolution during the phase transition, we calculate the charge compressibility,
\beq
\chi=\beta\chi_c=\frac{\beta}{N}\sum_{{\bf i},{\bf j}}\left[ \langle n_{\bf i} n_{\bf j}\rangle - \langle n_{\bf i}\rangle \langle n_{\bf j}\rangle \right]=\frac{d\langle n\rangle}{d\mu},\label{charge}
\eeq
which is measured directly in experiments\cite{YihangZeng,foutty2023mapping}. 
$\chi\rightarrow0$ when the system has a single-particle charge gap and hence insulating behavior. We first explore two topological phase transitions between QSH and trivial insulators, driven by $\lambda_v$ with $U=0$ and by $U$ with $\lambda_v=0$ respectively at half-filling $\langle n \rangle=2$. Without loss of generality, we set $t'/t=0.2$ and $\psi=0.5$ (the original KM model). The compressibility at the minimal finite field ($\Phi/\Phi_0=1/36\approx0.028$) and half-filling from DQMC simulation is plotted in \figdisp{fig:chi}(a) and (b) respectively to describe these two transitions at varying temperatures. Since both QSH and trivial insulator have a zero charge Chern number, they remain insulating at a small magnetic flux and $\langle n \rangle=2$ according to Streda's formula. Thus the compressibility at the minimal flux can well represent its zero-field counterpart while having the least finite size effect\cite{mfp}.

Consider first the non-interacting KM model studied in (\figdisp{fig:chi}(a)).  Here the system is a QSH insulator with a spin Chern number $C_s=2$ for small $\lambda_v$ and turns into a trivial band insulator at $\lambda_v=3\sqrt{3}t'\sin{\psi}\approx1.039$ accompanied by closing and reopening of the charge gap. The compressibility $\chi$ characterizing this transition in \figdisp{fig:chi}(a) has a peak at the transition point which only sharpens as the temperature is lowered.  Away from $\lambda_v=1.039$, the peak in the compressibility vanishes and insulating behaviour obtains.  This is the standard story.  In the interacting case (\figdisp{fig:chi}(b)), the QSH insulator converts into an $xy$-antiferromagnetic trivial Mott insulator as $U$ becomes sufficiently large ($U_c \approx 5.2$ labeled by the dashed line)\cite{Hohenadler2011,Hohenadler2012,WeiWu}. \figdisp{fig:chi}(b) is qualitatively different from \figdisp{fig:chi}(a). Although there is a soft peak around $U_c$ at $\beta=5$ suggesting the gap becomes smaller, at sufficiently low temperature $\beta=12$ so as to resolve the gap ($\beta\Delta\gg1$), the compressibility vanishes everywhere, meaning that the charge gap remains finite across the transition. This is not quite surprising since a non-interacting TPT with explicit symmetry breaking can also happen without closing the single-particle gap\cite{Nagaosa}. However, since this is a Mottness-driven TPT, a single-particle picture is not sufficient to capture the full interplay between correlation-driven magnetism and topological Mott insulation. A non-perturbative many-body theory is needed to understand this phenomenon. Note that although the charge gap remains open, the large-U antiferromagnetic ordering indicates gapless spin wave excitations, namely a closing of the many-body energy gap, as expected.  

More relevant to the TMD materials\cite{TingxinLi,YihangZeng,nogapxiaodong} is the TPT at quarter-filling. An earlier study \cite{quarter} demonstrated that a ferromagnetic QAH Mott insulator emerges at 1/4-filling with a flat lower band and sufficiently large $U$. This case must lie outside the paradigm of topological band insulators since the system is metallic at small $U$. Here we choose $t'=0.3$, $\psi=0.81$ which bears a non-interacting flatness (bandwidth to gap) ratio $r(\lambda_v=0)\approx1/6$ and keep a strong interaction strength of $U=8$. In analogy with the non-interacting case, turning on $\lambda_v$ lifts the inversion symmetry and drives the insulator to a trivial one. A QAH insulator is observed at $\lambda_v=0.2$ while a trivial paramagnetic insulator obtains at $\lambda_v=1$ at least for $U\geq3$ (see supplement). Thus, a TPT must occur between these values. For the TPT between QAH and trivial insulators, we have to study the zero-field 1/4-filling point which could suffer from finite size effects at low temperatures in DQMC. Also, the presence of a serious sign problem in DQMC prevents us from going to sufficiently low temperatures to resolve the gap which is much smaller than the half-filling one\cite{quarter}. Therefore, we use DCA \cite{MaierRMP2005,DCApp,Mai2,Mai3} on a $2\times2$, i.e. 4-unit cells, 8-site cluster with a continuous-time QMC\cite{GullEPL2008} cluster solver to calculate the zero-field compressibility. DCA treats the infinite lattice in the thermodynamic limit by restricting spatial correlations to a finite-size cluster that is embedded in a self-consistent dynamical mean field. Given the dynamical mean field, DCA has a milder sign problem and less finite-size errors compared to DQMC. The zero-field compressibility at quarter-filling from DCA simulations is shown in \figdisp{fig:chi}(c) for varying temperatures with $U=8$. \figdisp{fig:chi}(c) is qualitatively similar to \figdisp{fig:chi}(b). As the temperature is reduced, the compressibility vanishes for all $\lambda_v$, indicating the charge gap also remains finite during this transition, which is consistent with the experiment for $\nu=1$ AB stacked MoTe$_2$/WSe$_2$ heterobilayers\cite{TingxinLi}. In \figdisp{fig:chi}(d), the compressibility at the lowest temperature, $\beta=20$, develops a higher and sharper peak as $U$ decreases. This suggests that, for a smaller $U$, the transition could take place with a closure of the single-particle gap, which explains the experimental observation in $\nu=1$ twisted bilayer MoTe$_2$\cite{nogapxiaodong,YihangZeng} and WSe$_2$\cite{foutty2023mapping}. During the TPT, the gap is always suppressed to some extent indicated by the peak of the compressibility. If the topological gap is sufficiently large (due to a large $U$), it only becomes smaller but does not vanish at the transition. If the effective flatness ratio (which also controls the gap size) is tunable for example by the twist angle, such a change in gap evolution may also appear. 

In short, the TPT between topological and trivial insulators in the presence of sufficiently strong interactions at either half-filling (\figdisp{fig:chi}(b)) or quarter-filling (\figdisp{fig:chi}(c)) appears generally without closing the single-particle gap and without the introduction of explicit symmetry-breaking fields. This is contradictory to the traditional topological band theory, and indicates that these interaction-enabled topological phase transitions are intrinsically many-particle effects.

\section{Extended orbital HK interaction}

To decipher this enigma, it is useful to work with a solvable model that exhibits such a TPT. To this end, we adopt the HK\cite{HK,HKnp1,HKnp2,Haldanequarter,quarter,zhaoprl,barry,polishgroup} interaction which has been used extensively recently.  The drawback of this interaction, namely a thermodynamic degeneracy is completely lifted by resorting to the orbital HK interaction\cite{manning2023ground}.  In this model, it is the hybridization between the orbitals that lifts the degeneracy except at time-reversal invariant momenta. 
We will refer to the traditional same sublattice repulsion of the HK model as $U'$ and $V'$ as the nearest-neighbor inter-sublattice repulsion, both between opposite spins to represent Mottness. With this in mind, we write the KM model with extended orbital HK interaction as (at zero field $\phi_{{\bf i},{\bf j}}=0$),
\beq
\begin{aligned}
H_{\text{KM-HK}}&=H_{\text{KM}}+H_{\text{HK}},\\
H_{\text{HK}}&=U^\prime\sum_{\alpha{\bf k }} n_{\alpha{\bf k }\uparrow} n_{\alpha{\bf k }\downarrow} + V^\prime\sum_{{\bf k }\sigma}n_{A{\bf k }\sigma}n_{B{\bf k }\bar{\sigma}}.
\end{aligned} \label{KMHK1}
\eeq
We keep $t'=0.2$ and $\psi=0.5$ for illustration. As we will show below, with $V'>0$, there is a transition to a trivial Mott insulator without closing the single-particle gap given a sufficiently large $U'$.

This model is exactly solvable since the ${\bf k }$'s are independent. The model is further simplified if $U'=V'$. In this case, the Hamiltonian in \disp{KMHK1} has a simple form in the band basis,
\beq
\begin{aligned}
H_{\text{KM-HK}} &= \sum_{\gamma{\bf k }\sigma}\varepsilon_{\gamma{\bf k }\sigma}c^\dagger_{\gamma{\bf k }\sigma}c_{\gamma{\bf k }\sigma}- \mu\sum_{\gamma{\bf k }\sigma} n_{\gamma{\bf k }\sigma}
\\&+ U'\sum_{{\bf k }} (n_{+{\bf k }\uparrow}+n_{-{\bf k }\uparrow}) (n_{+{\bf k }\downarrow}+n_{-{\bf k }\downarrow}),  \label{UV}
\end{aligned} 
\eeq
where $\gamma=\pm$ represents the band index of the upper and lower bands from diagonalizing the non-interacting KM model. In the following, we will explore the physics at half-filling for general $U'$ and $V'$ with special attention to $U'=V'$. 

\begin{figure}[ht]
    \centering
    \includegraphics[width=0.5\textwidth]{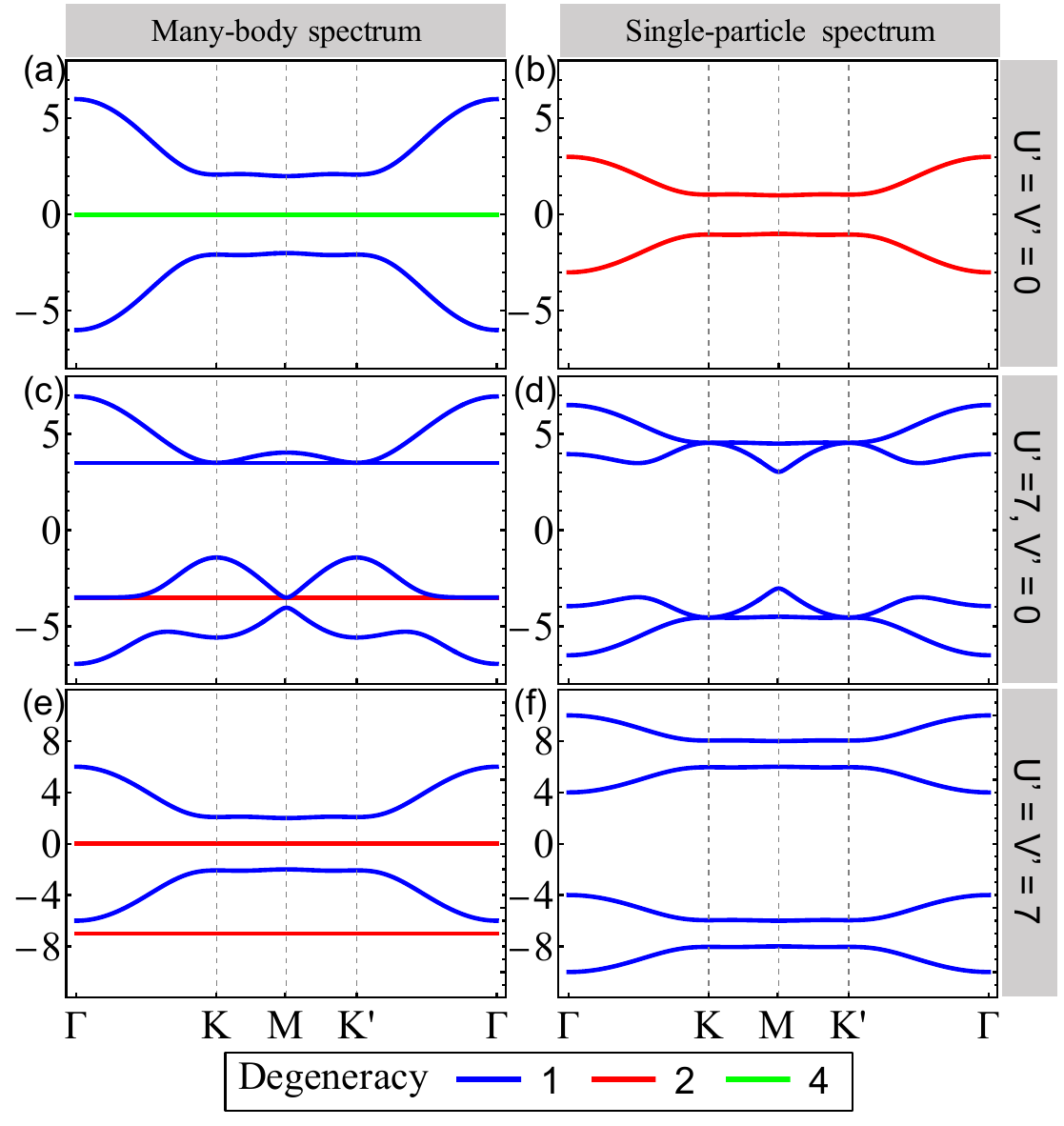}
    \caption{The many-body (left column) and single-particle (right column) spectra for the KM model with extended orbital HK interaction. Panels (a) and (b) show the non-interacting limit. The second (c-d) and third rows (e-f) display the cases at $U'/t=7, V'/t=0$ and $U'/t=V'/t=7$ respectively. The chemical potential is $\mu=U'/2+V'/2$.
    }
    \label{fig:two_spectrum}
\end{figure}

\begin{figure}[ht]
    \centering
    \includegraphics[width=0.5\textwidth]{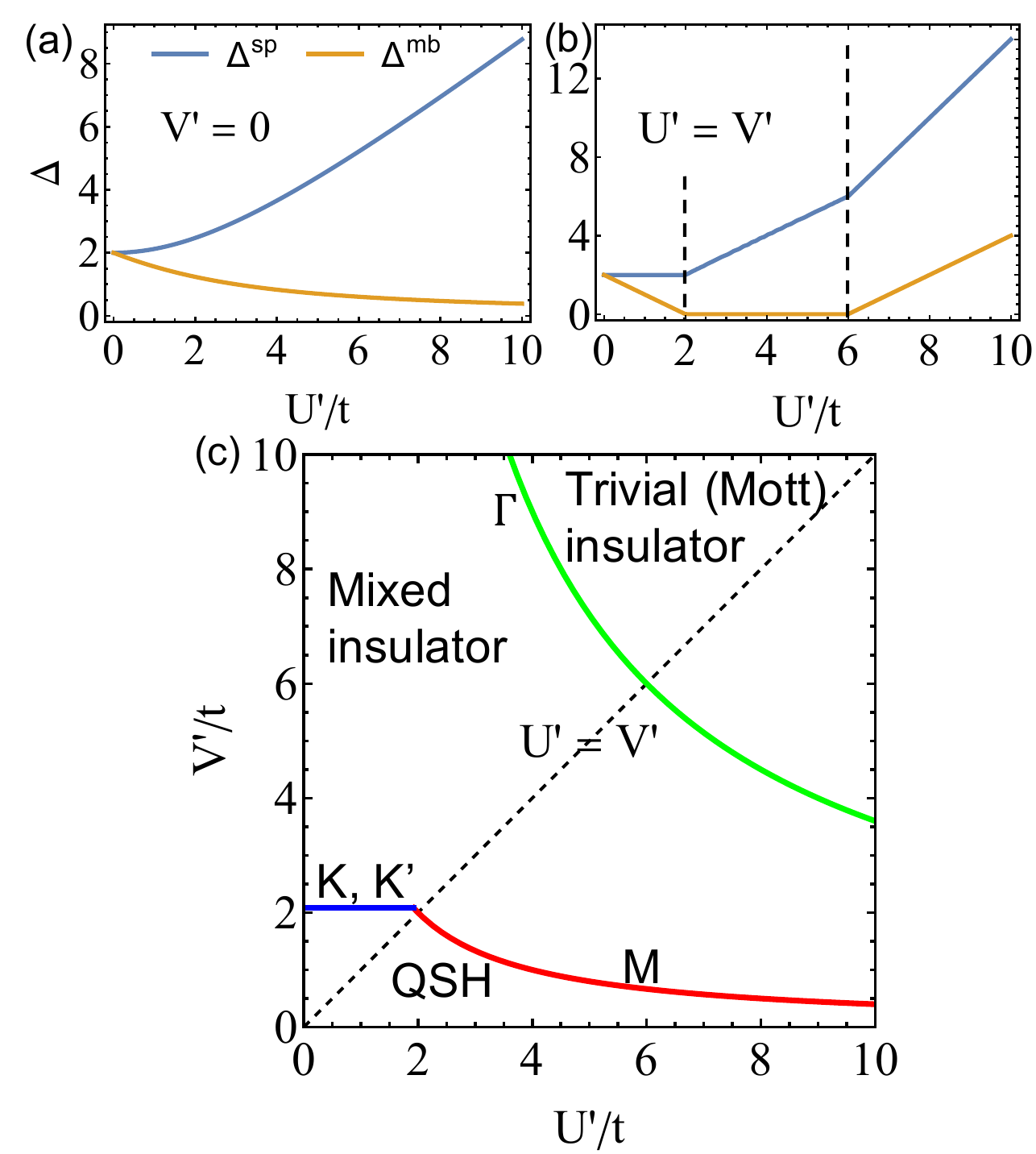}
    \caption{The single-particle and many-body gaps as well as the phase diagram at half-filling. Panel (a) plots the gap as a function of $U'$ at $V'=0$. Panel (b) shows the dependence on $U'$ if fixing $U'=V'$. The dashed lines are located where the many-body gap closes and therefore are indicators of the phase transition. Panel (c) shows the phase diagram of the KM model with extended orbital HK interaction.
    }
    \label{fig:two_gap}
\end{figure}

We first solve the KM-HK Hamiltonian to obtain the many-body and single-particle spectra with the chemical potential shifted to $\mu=U'/2+V'/2$. The many-body spectrum $\varepsilon^\text{mb}_{\eta,{\bf k}}$ at $\langle n_{\bf k}\rangle=2$ is plotted in \figdisp{fig:two_spectrum}(a,c,e) for $U'=V'=0$, $U'=7\& V'=0$ and $U'=V'=7$ respectively, obtained by diagonalizing the Hamiltonian at each ${\bf k}$. Here $\eta$ is the band index running from 1 to 6 since there are six states at each ${\bf k}$ for $n_k=2$, and $\eta=1$ represents the ground state. In a general quantum phase transition\cite{sachdev_2011}, there must be a level crossing between the many-body ground state and the first excited state. Thus, we focus on the lowest two states, which are usually the degenerate spin-triplet states ($|\uparrow\uparrow\rangle$ or $|\downarrow\downarrow\rangle$ labeled in red of \figdisp{fig:two_spectrum}(c,e)) and the topological state labeled in blue. With $U'$ and $V'$ nonzero, the QSH state gains energy, while the trivial singly-occupied spin-triplet state remains unchanged. Eventually, when $U'$ and $V'$ reach a threshold, a level crossing for TPT occurs and the spin-triplet state becomes the ground state. Hence, the level crossing in the many-body spectrum could be an indicator of the TPT. However, the many-body spectrum for the Hubbard model can only be obtained using exact diagonalization which is limited to only small clusters and is significantly influenced by finite size effects. 

Experimentally, however, it is the single-particle spectrum that is more accessible.  What we show now is that the single-particle spectrum is likely to be a poor indicator of the transition.   We obtain the single-particle spectrum defined as the ground state energy difference of adding a particle $\varepsilon^\text{sp}_{m,{\bf k}}=\varepsilon^\text{mb}_{1,{\bf k}}(\langle n_k\rangle= m)-\varepsilon^\text{mb}_{1,k}(\langle n_k\rangle= m-1)$ by computing the corresponding many-body spectrum to acquire the needed energy differences. The results are shown in \figdisp{fig:two_spectrum}(b,d,f) for various interaction strengths. There are four separated single-particle bands in \figdisp{fig:two_spectrum} (d) and (f) because interactions separate the doubly-occupied states from the singly-occupied ones. Unlike the non-interacting spinful counterpart in \figdisp{fig:two_spectrum}(a) with a degeneracy of 2, each interacting single-particle sub-band can only fill one particle for each momentum. At half-filling, the lowest two single-particle sub-bands are filled. Unlike the many-body spectrum, since all single-particle bands are nondegenerate (i.e., blue in Fig.~\ref{fig:two_gap}), it is not obvious to determine whether level crossing happens.


To further explore the possible level crossing in these spectra, we calculate the many-body gap $\Delta^{\text{mb}}$ which vanishes at the TPT, and the single-particle gap $\Delta^{\text{sp}}$ which is usually measured in transport experiments\cite{nogapxiaodong,YihangZeng}. At half-filling, they are
\beq
\begin{aligned}
&\Delta^{\text{mb}}(\langle n\rangle=2)=
\\&\ \max(\min(\varepsilon^\text{mb}_{2,{\bf k}}(\langle n_k\rangle= 2))-\max(\varepsilon^\text{mb}_{1,{\bf k}}(\langle n_k\rangle= 2)),0)
\end{aligned}
\eeq
and
\beq
\Delta^{\text{sp}}(\langle n\rangle=2) = \max( \min( \varepsilon^\text{sp}_{3,{\bf k}})-\max(\varepsilon^\text{sp}_{2,{\bf k}}),0),
\eeq
where $\Delta^{\text{mb}}$ represents the lowest energy cost for multiparticle excitations, such as particle-hole pairs creation, while $\Delta^{\text{sp}}$ represents the lowest energy cost for a single-particle excitation. 
As mentioned above, $\Delta^{\text{mb}}$ vanishes definitely when a TPT obtains. In the non-interacting case, $\Delta^{\text{mb}}(\langle n\rangle)=\Delta^{\text{sp}}(\langle n\rangle)$. Thus in topological band theory, we learn that a TPT is always accompanied by closing and reopening the single-particle gap. In interacting systems, however, the single-particle gap could depart from the many-body gap. \figdisp{fig:two_gap}(a,b) shows two such examples of the $U'$-dependence of both gaps. In \figdisp{fig:two_gap}(a), we fix $V'=0$ and increase $U$, since it is natural to expect intra-orbital repulsion to be dominant, corresponding to the orbital-HK model\cite{barry}. The single-particle gap increases monotonically while the many-body gap decreases monotonically and approaches zero asymptotically for $U'\rightarrow\infty$. Thus, no topological transition occurs during this process and the system remains a QSH insulator, but $\Delta^{\text{sp}}>\Delta^{\text{mb}}$ due to repulsive interactions. In \figdisp{fig:two_gap}(b), we find that increasing $U'$ to $V'=U'$ and that the many-body gap closes at $U'=2$ and reopens at $U'=6$ (both highlighted with dashed lines), while the single-particle gap remains finite all the entire time though exhibits kinks at those two interaction strengths.  This is similar to the finite dip of the single-particle gap in the Hubbard case\cite{WeiWu}. In this process, the topology changes from a QSH state ($U'<2$) to a trivial Mott insulator ($U'>6$).

\begin{figure}[ht]
    \centering
    \includegraphics[width=0.5\textwidth]{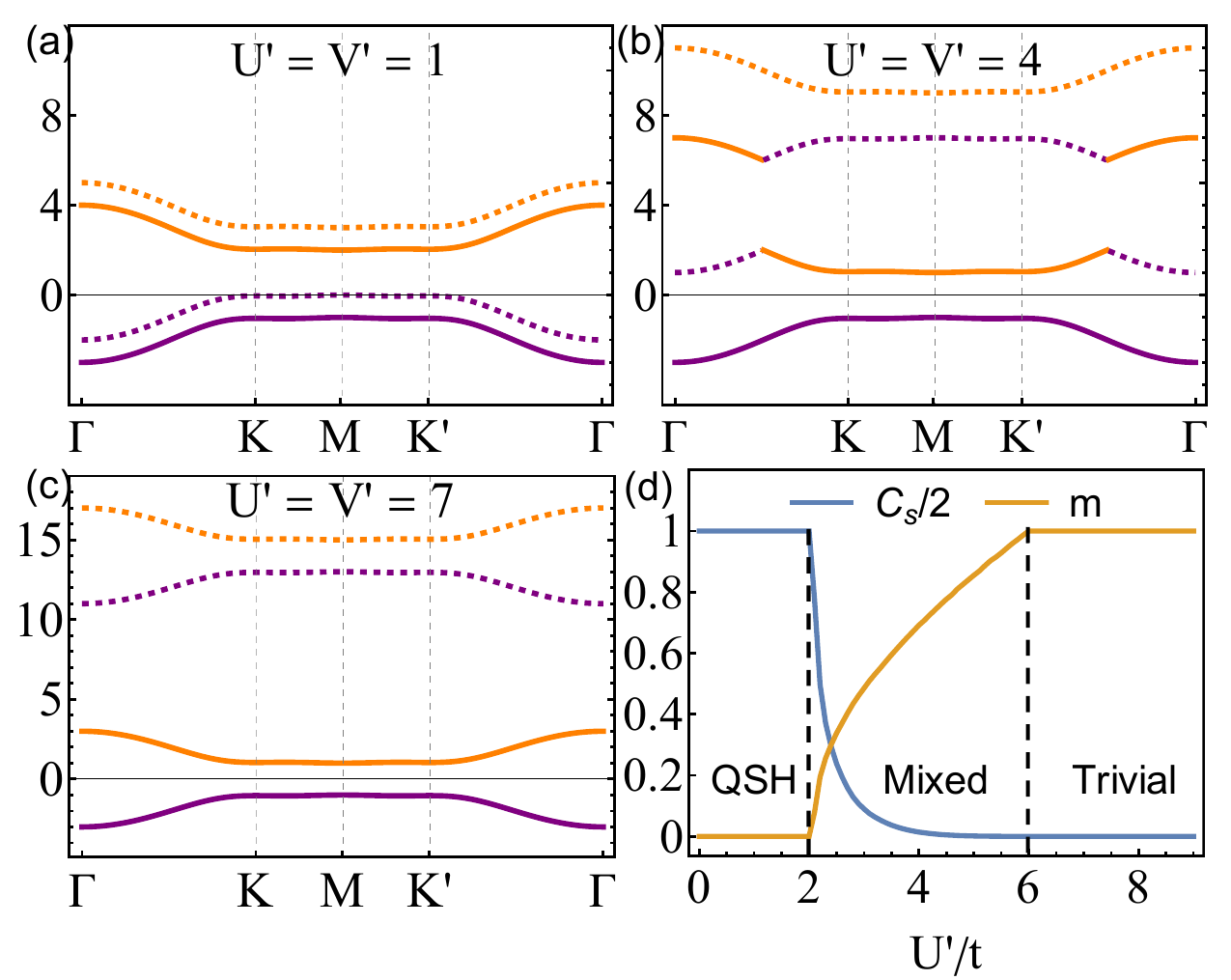}
    \caption{The single-particle spectra are presented for (a) QSH insulator, (b) mixed insulator, and (c) trivial Mott insulator. Purple and orange color represent  $C_s=1$ and $C_s=-1$ sub-bands, respectively. Singly and doubly occupied sub-bands are denoted by solid and dashed lines, respectively. The evolution of the spin Chern number and spontaneous magnetization with $U'$ is shown in panel (d). In all cases, we set $U'=V'$ and fix the chemical potential to 0.
    }
    \label{fig:spin_Chern}
\end{figure}

By analytically calculating where the many-body gap closes (see supplement), we obtain the phase diagram in \figdisp{fig:two_gap}(c). The blue, red and green lines depict the phase boundaries representing level-crossing of the many-body spectrum at K$\&$K$^\prime$ ($V'=6\sqrt{3} t'\approx2.078$), M ($U'V'=4$) and $\Gamma$ ($U'V'=36$) points respectively. We observe three different phases. The system is a QSH insulator with weak interactions and becomes a trivial Mott insulator for adequately strong $U'$ and $V'$. The intermediate state is an insulator with a finite single-particle gap (except at a single point $U'=0,V'=6\sqrt{3}t'$). We call this a mixed insulator because it is composed of the QSH state and the trivial triplet state in different parts of the Brillouin zone, which we further elaborate in \figdisp{fig:spin_Chern}.

From the phase diagram in \figdisp{fig:two_gap}(c), the dashed line $U'=V'$ crosses all three phases and hence offers a clear view of the phase transition. As mentioned earlier, the Hamiltonian becomes particularly simple (\disp{UV}) in this case because the eigenstates can be constructed by rearranging the non-interacting eigenstates with different eigenvalues. We plot the single-particle spectrum for $U'=V'=1, 4, 7$ in \figdisp{fig:spin_Chern}(a-c) respectively. In this case, we use a different color scheme from \figdisp{fig:two_spectrum} since all sub-bands here are non-degenerate;  further, we fix the chemical potential $\mu=0$ for convenient comparison. We use purple for $C_s=1$ and orange for $C_s=-1$ sub-bands. As $U'$ turns on, the doubly-occupied states (dashed curve) elevate, leaving the singly-occupied (solid curve) states in the lower band unchanged. Note that the singly-occupied states in the upper band also rise in energy due to the inter-band repulsion from the purple dashed band. For $U'=V'=1$ in \figdisp{fig:spin_Chern}(a), half-filling still corresponds to fully occupying the lower two (purple solid and dashed) sub-bands, indicating a QSH insulator as in the non-interacting case. When the energy of the lower doubly occupied (purple-dashed) sub-band increases to touch the original (non-interacting) upper band (at $U'=2$), the upper singly occupied (orange-solid) sub-band starts to replace the lower doubly occupied (purple-dashed) sub-band gradually over the Brillouin zone while a single-particle gap is maintained by the inter-band repulsion. This process occurs for $2<U'<6$ (\figdisp{fig:spin_Chern}(b)) in which case the second lowest band contains the lower doubly occupied and upper singly occupied states in different parts of the Brillouin zone. For this reason, we named this intermediate state a mixed insulator. Once the replacement is complete at $U'>6$ (\figdisp{fig:spin_Chern}(c)), the half-filling system becomes a trivial insulator with the two lowest sub-bands occupied yielding $C_s=0$. 
We deduce that the transition is driven by the intra-band repulsion, while it is the inter-band repulsion that maintains a single-particle gap throughout the transition. For Hubbard or other more realistic interactions, the interaction term does not commute with the kinetic term. This always leads to an inherent intra- and inter-band contribution if written in the band basis. Thus, the single-particle gap is allowed to remain open during the TPT in the presence of interactions. This explains our previous observation on a persistent single-particle gap in the TPT for the Hubbard model.

We also calculate the spin Chern number\cite{quarter} and magnetization (see supplement) as a function of $U'$, shown in \figdisp{fig:spin_Chern}(d). As expected, $C_s=2$ for the QSH state at $U<2$ and $C_s=0$ for the trivial state at $U>6$. 
In the intermediate region, since the many-body gap remains closed, the Chern number is not well-defined, namely, not quantized, as if the system were in a metallic state in the sense of band theory. But here it is a correlated insulator with an emergent charge gap due to the interactions. It still remains valid to define the spin Chern number using the summation of the Berry curvature over the filled mixed states in the Brillouin zone. Thus, we see a continuous change of spin Chern number for $U'$ increasing from $2$ to $6$. The degenerate singly-occupied spin-triplet states (at each ${\bf k}$) are unstable against spontaneous symmetry breaking towards ferromagnetism at zero temperature, while the QSH state is paramagnetic. Thus, by counting the number of ${\bf k}$ points in which the triplet state is the ground state, we obtain the spontaneous magnetization $m$. As shown in \figdisp{fig:spin_Chern}(d), it turns on at $U'=2$ and subsequently increases until $U'=6$ at which point it saturates. This phase transition is continuous according to the evolution of the magnetization. The solvable HK model provides a clear example at half-filling to elucidate how a TPT between a topological and trivial insulator obtains without closing the single-particle gap. The HK model at quarter-filling with a non-zero $\lambda_v$ can exhibit strange topology, where spin-up particles occupy one half of the Brillouin zone and spin-down particles the other half. 

\section{Discussion}

Although the trivial insulating states differ in the Hubbard (antiferromagnet) and HK (ferromagnet) models, both interactions containing Mottness show a TPT between a topological and trivial insulator without closing the single-particle gap for the half-filled interacting KM model. This indicates the generality of this emergent phenomenon as a direct cause of topological Mottness. Sufficiently strong interactions ($U>\Delta_0$ and $U>W$) are needed to observe this in experiments. The transition between QSH and trivial band insulator in AB stacked MoTe$_2$/WSe$_2$ moir\'{e} heterobilayers at $\nu=2$\cite{TingxinLi} is essentially the non-interacting transition as in \figdisp{fig:chi}(a) and hence the charge gap closes and reopens when the topology changes. At $\nu=1$ of the same material\cite{TingxinLi}, corresponding to $1/4-$filling of our model, a topological phase transition between QAH and trivial Mott insulators is observed without a single-particle gap-closing. However, this phenomenon is likely to not be general but specific to this material because a similar transition has been observed in $\nu=1$ twisted bilayer MoTe$_2$\cite{JiaqiCai,YihangZeng} with a closing of the single-particle gap. In our DCA simulation, depending on the Hubbard interaction strength, the compressibility plots at $1/4-$filling support both gap-not-closing and gap-closing topological phase transitions, as is observed differently in these two materials.  Consequently, which scenario wins is likely to be controlled by the details of the interaction strength and hence the flatness of the bands.  Nonetheless, our work here offers examples of both scenarios and represents a platform to understand the range of experimental results.

We also emphasize that the gap-not-closing TPTs in our models occur in the absence of any external symmetry-breaking fields. For all values of the interaction strength and tuning parameters, the symmetries of the Hamiltonian do not change. Nevertheless, the TPTs with no single-particle gap closing in our models are accompanied by the onset of spontaneous symmetry breaking in the ground state. The gapless Goldstone modes associated with the spontaneous symmetry breaking emerge via the closure of the many-body energy gap, and ensure that the TPT occurs when the many body gap closes. Our work thus gives a completely microscopic picture of correlation-driven TPTs in electronic systems beyond mean-field theory, without the need to introduce explicit symmetry-breaking fields. Our findings are also relevant to the recent advancement in the field of interacting topological quantum chemistry \cite{Bernevig2022}.

\textbf{Acknowledgements} 
We thank Kai Sun, Edwin W. Huang, Kin Fai Mak, Yihang Zeng and Charlie Kane for useful discussions. This work was supported by the Center for Quantum Sensing and Quantum Materials, a DOE Energy Frontier Research Center, grant DE-SC0021238 (P.M., B.B., and P.W.P.). PWP also acknowledges NSF DMR-2111379 for partial funding of the HK work which led to these results. The analytical work of B.B. on orbital HK models was partially supported by the Alfred P Sloan foundation and the National Science Foundation under grant DMR-1945058. The contributions of T.A.M. to the DCA calculations were supported by the U.S. Department of Energy, Office of Science, Basic Energy Sciences, Materials Sciences and Engineering Division. The DQMC calculations used the Advanced Cyberinfrastructure Coordination Ecosystem: Services \& Support (ACCESS) Expanse supercomputer through the research allocation TG-PHY220042, which is supported by National Science Foundation grant number ACI-1548562\cite{xsede}. The DCA calculations were supported through the INCITE program and used resources of the Oak Ridge Leadership Computing Facility, which is a DOE Office of Science User Facility supported under Contract No. DE-AC05-00OR22725.
\bibliography{reference}

\end{document}


\title{Topological Phase Transition without Single-Particle-Gap Closing in Strongly Correlated Systems: supplementary information}
\author{ Peizhi Mai$^{1}$, Jinchao Zhao$^{1}$, Thomas A. Maier$^{2}$, Barry Bradlyn$^{1}$
 and Philip W. Phillips$^{1,\dagger}$ }

\affiliation{$^1$Department of Physics and Institute of Condensed Matter Theory, University of Illinois at Urbana-Champaign, Urbana, IL 61801, USA}
\affiliation{$^2$Computational Sciences and Engineering Division, Oak Ridge National Laboratory, Oak Ridge, Tennessee 37831, USA}


\begin{abstract}

\end{abstract}
\date{October 2022}


\maketitle
\section{Different topological phases at quarter-filling of the generalized Kane-Mele-Hubbard model}
\begin{figure*}[ht]
    \centering
    \includegraphics[width=0.7\textwidth]{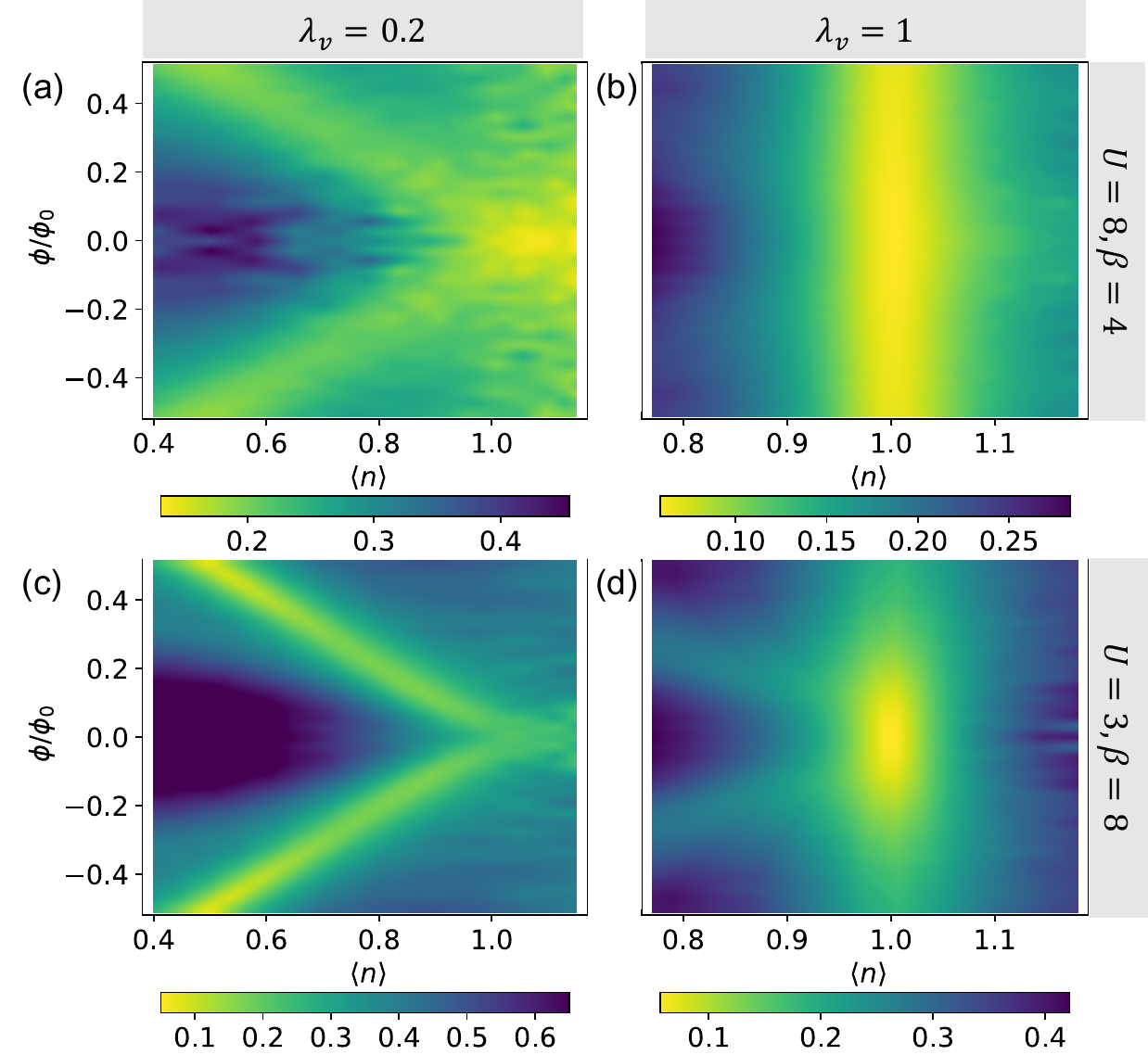}
    \caption{The compressibility is plotted as a function of magnetic flux and density for (a) $U/t=8, \beta=4/t, \lambda_v/t=0.2$; (b) $U/t=8, \beta=4/t, \lambda_v/t=1$; (c) $U/t=3, \beta=8/t, \lambda_v/t=0.2$ and (d) $U/t=3, \beta=8/t, \lambda_v/t=1$. The other parameters are $\lambda_{\text{SO}}=0.3, \psi=0.41$ (in the unit of $\pi$).
    }
    \label{fig:sup_topo}
\end{figure*}

In this section, we show different topological phases at quarter-filling of the generalized Kane-Mele-Hubbard (KMH) model. Given a relatively flat lower band, a gap emerges at quarter-filling in the presence of sufficiently strong interaction. To determine the topological nature of the gap, we look into the compressibility as a function of magnetic flux and particle density, shown in \figdisp{fig:sup_topo}. According to Streda's formula, we find a QAH insulator ($C=\pm1$) at $\lambda_v=0.2$ (\figdisp{fig:sup_topo}(a,c)) and a trivial insulator ($C=0$) at $\lambda_v=1$ (\figdisp{fig:sup_topo}(b,d)) for both $U=3$ and $U=8$. Thus, a topological phase transition must occur between $\lambda_v=0.2$ and $\lambda_v=1$ for at least $U\geq3$. 

\section{Calulation of phase boundary at half-filling for the extended orbital HK model}
In this section, we present the details of the analytical calculation to obtain the phase boundary in Fig.~3 of the main text. From plotting the many-body spectrum at different $U$ and $V$, we notice that the first and last band touching between the lowest two bands always happens at the high symmetry points: $\Gamma, M$ and $K$ ($K'$). Thus, we can obtain the phase boundaries by working out the conditions for band touching at those points. 

Let's rewrite the Hamiltonian in momentum space:
\beq
\begin{aligned}
H &= t\sum_{{\bf k }\sigma}g_0({\bf k })c^\dagger_{A{\bf k }\sigma}c_{B{\bf k }\sigma}+ h.c.
\\&\ + i\lambda_{\text{SO}}\sum_{\alpha{\bf k }\sigma} \alpha\sigma g_1({\bf k }) c^\dagger_{\alpha{\bf k }\sigma}c_{\alpha{\bf k }\sigma}  + h.c.- \mu\sum_{\alpha{\bf k }\sigma} n_{\alpha{\bf k }\sigma}
\\&+ U\sum_{\alpha{\bf k }} n_{\alpha{\bf k }\uparrow} n_{\alpha{\bf k }\downarrow} + V\sum_{{\bf k }\sigma}n_{A{\bf k }\sigma}n_{B{\bf k }\bar{\sigma}} \label{KMHK}
\end{aligned}
\eeq
At half-filling, namely the $N=2$ sector, there are six states. We work with the following ordered basis (ignoring ${\bf k}$ index):
\beq
\begin{aligned}
\big|11\rangle &= c^\dagger_{A\uparrow}c^\dagger_{B\uparrow}\big|0\rangle
\\ \big|1\bar{1}\rangle &= c^\dagger_{A\downarrow}c^\dagger_{B\downarrow}\big|0\rangle
\\ \big|10\rangle &= \frac{1}{\sqrt{2}}(c^\dagger_{A\uparrow}c^\dagger_{B\downarrow}+c^\dagger_{A\downarrow}c^\dagger_{B\uparrow})\big|0\rangle
\\ \big|a\rangle &= \frac{i}{\sqrt{2}} (c^\dagger_{A\uparrow}c^\dagger_{A\downarrow}+c^\dagger_{B\uparrow}c^\dagger_{B\downarrow})\big|0\rangle
\\ \big|b\rangle &= \frac{1}{\sqrt{2}} (c^\dagger_{B\uparrow}c^\dagger_{B\downarrow}-c^\dagger_{A\uparrow}c^\dagger_{A\downarrow})\big|0\rangle
\\ \big|c\rangle &= \frac{1}{\sqrt{2}} (c^\dagger_{A\uparrow}c^\dagger_{B\downarrow}-c^\dagger_{A\downarrow}c^\dagger_{B\uparrow})\big|0\rangle.
\end{aligned}
\eeq
The three states $\{\big|11\rangle, \big|1\bar{1}\rangle, \big|10\rangle \}$ have total spin $S=1$, while the other three states $\{\big|a\rangle, \big|b\rangle, \big|c\rangle \}$ are spin singlet states with total spin $S=0$. Now we write the Hamiltonian in this basis, denote as $H_2$:
\beq
H_2 = 
\begin{pmatrix}
0 & 0 & 0 & 0 & 0 & 0 \\
0 & 0 & 0 & 0 & 0 & 0 \\
0 & 0 & V & 0 & 0 & 2 d_z \\
0 & 0 & 0 & U & 0 & 2i d_x \\
0 & 0 & 0 & 0 & U & 2i d_y \\
0 & 0 & 2 d_z & -2i d_x & -2i d_y & V 
\end{pmatrix}
-2\mu \label{matrix}
\eeq
where $d_x = \Re(g_0), d_y = -\Im(g_0), d_z = -2\Im(g_1)$ (all are functions of ${\bf k}$) The term $-2\mu$ can be dropped for the calculation of band touching as it shifts all bands equally. From \disp{matrix}, the relevant matrix dimension is $4\times4$.

When ${\bf k}=\Gamma=(0,0)$, then $d_x=3, d_y=0, d_z=0$. The $4\times4$ matrix becomes 
\beq
H_2'(\Gamma) = 
\begin{pmatrix}
V & 0 & 0 & 0 \\
0 & U & 0 & 6i \\
0 & 0 & U & 0 \\
0 & -6i & 0 & V 
\end{pmatrix}
\label{matrix2}
\eeq
There are two eigenvalues $U$ and $V$ by observation. Band overlap starts to happen when the lowest energy of the above matrix becomes zero, the same as the first two spin-triplet states. Plugging that into the eigenvalue equation $(U-x)(V-x)-36=0$, we obtain the condition $UV=36$.

When ${\bf k}=K=(2\pi/3,2\pi/(3\sqrt{3}))$, then $d_x=0, d_y=0, d_z=-3\sqrt{3}\lambda_{\text{SO}}$. The $4\times4$ matrix becomes 
\beq
H_2'(K) = 
\begin{pmatrix}
V & 0 & 0 & -6\sqrt{3}\lambda_{\text{SO}} \\
0 & U & 0 & 0 \\
0 & 0 & U & 0 \\
-6\sqrt{3}\lambda_{\text{SO}} & 0 & 0 & V 
\end{pmatrix}
\label{matrix2}
\eeq
Similarly, a two-fold degenerate state with energy $U$ is obtained. Band overlap starts to happen when the lowest energy of the above matrix becomes zero, the same as the first two spin-triplet states. Plugging that into the eigenvalue equation $(V-x)^2-(6\sqrt{3}\lambda_{\text{SO}})^2=0$, we obtain the condition $V=6\sqrt{3}\lambda_{\text{SO}}$. When ${\bf k}=K'=(0,4\pi/(3\sqrt{3}))$, $d_x=0, d_y=0, d_z=3\sqrt{3}\lambda_{\text{SO}}$. Therefore the band-crossing condition is the same, as expected from the inversion symmetry of the model.

When ${\bf k}=M=(\pi/3,\pi/\sqrt{3})$, $d_x=1/2, d_y=\sqrt{3}/2, d_z=0$. The $4\times4$ matrix becomes 
\beq
H_2'(M) = 
\begin{pmatrix}
V & 0 & 0 & 0 \\
0 & U & 0 & i \\
0 & 0 & U & \sqrt{3}i \\
0 & -i & -\sqrt{3}i & V 
\end{pmatrix}
\label{matrix2}
\eeq
The eigenvalue equation for the lower $3\times3$ matrix is $(U-x)^2(V-x)-4(U-x)=0$. To obtain a solution $x=0$, we need the condition $UV=4$. In summary, we obtain all three phase boundaries for the phase diagram through analytical calculation.

\section{Calulation of spin susceptibility at half-filling}

\subsection{Hamiltonian}

The Hamiltonian is 
\beq
H = t\sum_{{\bf k}\sigma} g({\bf k}) c^\dagger_{A{\bf k}\sigma} c_{B{\bf k}\sigma} + i\lambda \sum_{\alpha{\bf k}\sigma}\sigma\alpha g_1({\bf k})n_{\alpha{\bf k}\sigma}-\mu\sum_{\alpha{\bf k}\sigma} n_{\alpha{\bf k}\sigma}+U\sum_{\alpha{\bf k}} n_{\alpha{\bf k}\uparrow} n_{\alpha{\bf k}\downarrow} + V\sum_{{\bf k}\sigma} n_{A{\bf k}\sigma} n_{B{\bf k}\bar{\sigma}}
\eeq
where $\alpha=1$ for A sublattice and $-1$ for B sublattice and
\beq
n_{\alpha{\bf k}\sigma} = c^\dagger_{\alpha{\bf k}\sigma} c_{\alpha{\bf k}\sigma}
\eeq
In the case $U=V$, it becomes particularly simple as 
\beq
H = \sum_{\gamma{\bf k}\sigma} (\varepsilon_{\gamma{\bf k}}-\mu) n_{\gamma{\bf k}\sigma} + U \sum_{{\bf k}} (n_{+{\bf k}\uparrow}+n_{-{\bf k}\uparrow} ) (n_{+{\bf k}\downarrow}+n_{-{\bf k}\downarrow} ) - \frac{h}{2}\sum_{\gamma{\bf k}\sigma} \sigma n_{\gamma{\bf k}\sigma}  \label{SH}
\eeq
where $\gamma=\pm$ is the band index and we add a small Zeeman field which would be sent to 0 at the end. 
\subsection{Partition function at half-filling}
With \disp{SH}, we can write down the partition function as 
\beq
Z = \prod_{{\bf k}} Z_{{\bf k}}
\eeq
For each ${\bf k}$, we have
\beq
Z_{{\bf k}} = Z_{0{\bf k}} + Z_{1{\bf k}} + Z_{2{\bf k}} + Z_{3{\bf k}} + Z_{4{\bf k}}
\eeq
where the number in subscript tells how many states are occupied at a given ${\bf k}$. First of all, $Z_{0{\bf k}} = 1$ and $Z_{4{\bf k}}=\exp(-\beta (4U-4\mu))$.

The others are
\beq
Z_{1{\bf k}} = \sum_{\gamma\sigma}\exp(-\beta (\varepsilon_{\gamma{\bf k}}-\mu-\frac{h\sigma}{2}))
\eeq
\beq
Z_{2{\bf k}} = 2\exp(-\beta (U-2\mu)) + \sum_{\sigma}\exp(-\beta (-2\mu-h\sigma)) + \sum_{\gamma}\exp(-\beta (2\varepsilon_{\gamma{\bf k}}+U-2\mu))
\eeq
Note that $\varepsilon_{+{\bf k}}=-\varepsilon_{-{\bf k}}$.
\beq
Z_{3{\bf k}} = \sum_{\gamma\sigma}\exp(-\beta (\varepsilon_{\gamma{\bf k}\sigma}+2U-3\mu-\frac{h\sigma}{2}))
\eeq
For half-filling, $\mu=U$, then $Z_{0{\bf k}} = Z_{4{\bf k}}=1$ and $Z_{1{\bf k}}=Z_{3{\bf k}}=\sum_{\gamma\sigma}\exp(-\beta (\varepsilon_{\gamma{\bf k}}-U-\frac{h\sigma}{2}))$. Therefore,
\beq
Z_{{\bf k}} = 2 + 2\sum_{\gamma\sigma}\exp(-\beta (\varepsilon_{\gamma{\bf k}}-U-\frac{h\sigma}{2})) + 2\exp(\beta U) + \sum_{\sigma}\exp(\beta (2U+h\sigma)) + \sum_{\gamma}\exp(-\beta (2\varepsilon_{\gamma{\bf k}}-U)) \label{pf}
\eeq

\subsection{Magnetization and magnetic susceptibility at half-filling}
With the partition function, we can calculate the magnetization as 
\beq
\begin{aligned}
& \langle m\rangle =\frac{1}{N} \sum_{{\bf k}} \langle m_{\bf k} \rangle
\\&\ \langle m_{\bf k} \rangle = \frac{1}{2}\sum_\gamma (\langle n_{\gamma{\bf k}\uparrow}\rangle-\langle n_{\gamma{\bf k}\downarrow} \rangle) =\frac{1}{\beta} \frac{\partial\ln Z}{\partial h} = \frac{1}{Z_{{\bf k}}}(\sum_{\gamma\sigma}\sigma\exp[\beta (-\varepsilon_{\gamma{\bf k}}+U+\frac{h\sigma}{2})]+\sum_{\sigma}\sigma\exp[\beta (2U+h\sigma)])
\end{aligned}
\eeq
Due to the factor of $\sigma$, $\langle m_{\bf k} \rangle\rightarrow 0$ if $h\rightarrow 0$. The magnetic susceptibility is 
\beq
\chi_s=\frac{d\langle m \rangle}{dh}\big|_{h\rightarrow0} = \frac{1}{N} \sum_{{\bf k}} \frac{d\langle m_{\bf k} \rangle}{dh}\big|_{h\rightarrow0} = \frac{1}{N} \sum_{{\bf k}} \chi_{s,{\bf k}}
\eeq
For a certain ${\bf k}$, 
\beq
\begin{aligned}
\chi_{s,{\bf k}} &= \frac{\beta}{Z_{{\bf k}}}(\exp[\beta (-\varepsilon_{+{\bf k}}+U)]+\exp[\beta (-\varepsilon_{-{\bf k}}+U)]+2\exp[2\beta U])
\\&\ = \frac{\beta}{Z_{{\bf k}}}\exp(\beta U) (\exp[-\beta \varepsilon_{+{\bf k}}] + \exp[\beta \varepsilon_{+{\bf k}}] + 2\exp[\beta U])
\\&\ = \frac{\beta\exp(\beta U)(\exp[-\beta \varepsilon_{+{\bf k}}] + \exp[\beta \varepsilon_{+{\bf k}}] + 2\exp[\beta U])}{2 + 2\sum_{\gamma\sigma}\exp[-\beta (\varepsilon_{\gamma{\bf k}}-U)] + 2\exp(\beta U) + 2\exp(2\beta U) + \sum_{\gamma}\exp(-\beta (2\varepsilon_{\gamma{\bf k}}-U)) }
\end{aligned}
\eeq
Here we use $\varepsilon_{+{\bf k}} = -\varepsilon_{-{\bf k}}$. With $\varepsilon_{+{\bf k}}>0$ in mind, taking the zero temperature limit $T\rightarrow0$ or $\beta\rightarrow\infty$, 
we have
\beq 
\chi_{s,{\bf k}}\big|_{\beta\rightarrow \infty} = \frac{\beta\exp(\beta U)(\exp(\beta \varepsilon_{+{\bf k}}) + 2\exp[\beta U])}{2\exp(2\beta U) + \exp[\beta (2\varepsilon_{+{\bf k}}+U)]}
\eeq
There are three situations: $U<\varepsilon_{+{\bf k}}$, $\varepsilon_{+{\bf k}}<U<2\varepsilon_{+{\bf k}}$ and $U>2\varepsilon_{+{\bf k}}$.

For $U<\varepsilon_{+{\bf k}}$, 
\beq
\chi_{s,{\bf k}}\big|_{\beta\rightarrow \infty} = \frac{\beta\exp(\beta U)\exp(\beta \varepsilon_{+{\bf k}}) }{\exp[\beta (2\varepsilon_{+{\bf k}}+U)]}=\beta\exp(-\beta\varepsilon_{+{\bf k}})\rightarrow 0.
\eeq
For $\varepsilon_{+{\bf k}}<U<2\varepsilon_{+{\bf k}}$, 
\beq
\chi_{s,{\bf k}}\big|_{\beta\rightarrow \infty} = \frac{2\beta\exp(2\beta U)}{\exp[\beta (2\varepsilon_{+{\bf k}}+U)]}=2\beta\exp[\beta(U-2\varepsilon_{+{\bf k}})]\rightarrow 0.
\eeq
For $U>2\varepsilon_{+{\bf k}}$, 
\beq
\chi_{s,{\bf k}}\big|_{\beta\rightarrow \infty} = \frac{2\beta\exp(2\beta U)}{2\exp(2\beta U)} = \beta.
\eeq
To summarize, approaching zero temperature, if $U<2\varepsilon_{+{\bf k}}$, $\chi_{s,{\bf k}}\rightarrow 0$ and if $U>2\varepsilon_{+{\bf k}}$, $\chi_{s,{\bf k}}\rightarrow \beta$. Since $1<\varepsilon_{+{\bf k}}<3$ ($2<2\varepsilon_{+{\bf k}}<6$) over the Brillouin zone, as shown in the Fig.~1 of the main text, then $\chi_s\rightarrow 0$ for $U<2$; $\chi_s\rightarrow p\beta $ for $2<U<6$ where $p$ is the proportion of the Brillouin zone with $2\varepsilon_{+{\bf k}}<U$; $\chi_s\rightarrow \beta$ for $U>6$.

The susceptibility calculated above is for magnetization along $z$ direction, namely $\chi_{sz,{\bf k}}$. We also calculate the susceptibility for magnetization along $x$ direction:
\beq
\begin{aligned}
\chi_{sx,{\bf k}}&= \frac{\beta}{4}\langle (\sum_{\gamma,\sigma}c^\dagger_{\gamma{\bf k}\sigma}c_{\gamma{\bf k}\bar{\sigma}})(\sum_{\gamma',\sigma'}c^\dagger_{\gamma'{\bf k}\sigma'}c_{\gamma'{\bf k}\bar{\sigma}'})\rangle
\\&\ =\frac{\beta}{Z_{{\bf k}}}(\exp[\beta (-\varepsilon_{+{\bf k}}+U)]+\exp[\beta (-\varepsilon_{-{\bf k}}+U)]+\exp(2\beta U)+\exp(\beta U))
\end{aligned}
\eeq
Taking the zero temperature limit $\beta\rightarrow \infty$, $\exp(2\beta U) \gg \exp(\beta U)$. Then $\chi_{sx,{\bf k}}\rightarrow \chi_{sz,{\bf k}}/2$
\bibliography{reference}